\documentclass[journal]{IEEEtran}
\usepackage{graphicx}
\usepackage{subfigure}
\usepackage{multirow}
\usepackage{array}
\usepackage{footnote}

\usepackage[ruled,linesnumbered]{algorithm2e}
\usepackage{dsfont}
\usepackage{amsfonts,amsmath}
\usepackage{url,cite}
\usepackage{color}
\usepackage{bm}
\usepackage{algorithmic}
\usepackage{booktabs}
 \usepackage{multirow}
 \usepackage[normalem]{ulem}
 \useunder{\uline}{\ul}{}

\begin{document}

\title{Unrolling Plug-and-Play Network for Hyperspectral Unmixing}

\author{Min~Zhao,~\IEEEmembership{IEEE Student Member,}
Linruize~Tang,~\IEEEmembership{IEEE Student Member,}
      and Jie~Chen,~\IEEEmembership{IEEE Senior Member}
       \vspace{-5mm}
\thanks{The authors are with School of Marine Science and Technology,
Northwestern Polytechnical University, China. (corresponding author: J.
Chen, dr.jie.chen@ieee.org).} }
\maketitle
\begin{abstract}
Deep learning based unmixing methods have received great attention in recent years and achieve remarkable performance. These methods employ a data-driven approach to extract structure features from hyperspectral image, however, they tend to be less physical interpretable. Conventional unmixing methods are with much more interpretability, whereas they require manually designing regularization and choosing penalty parameters. To overcome these limitations, we propose a novel unmixing method by unrolling the plug-and-play unmixing algorithm to conduct the deep architecture.
Our method integrates both inner and outer priors. The carefully designed unfolding deep architecture is used to learn the spectral and spatial information from the hyperspectral image, which we refer to as inner priors. Additionally, our approach incorporates deep denoisers that have been pretrained on a large volume of image data to leverage the outer priors.
Secondly, we design a dynamic convolution to model the multiscale information. Different scales are fused using an attention module. Experimental results of both synthetic and real datasets demonstrate that our method outperforms compared methods.
\end{abstract}


\IEEEpeerreviewmaketitle
\section{Introduction}
\label{sec:intro} \IEEEPARstart{H}yperspectral imaging stands as a pivotal domain in remote sensing, capturing data that incorporates both spatial and spectral characteristics of the target. The dense spectral information with hundreds of spectral bands allows it to be widely applied in many fields, such as environmental monitoring and agriculture~\cite{li2019deep,lu2020recent,briottet2006military}.
However, due to the high work distance and low spatial resolution of hyperspectral sensors, a pixel may contain several materials, which degrades the performance of subsequent high-level data processing. Spectral unmixing is one of the most prominent tools to cope with this issue, aiming to decompose the mixed pixels into pure components, termed as endmembers, and their corresponding
fraction abundances.

From the standpoint of physical interpretability and simplicity, the most commonly used mixing model is the linear mixing model (LMM), presuming that an observed pixel is a linear combination of endmembers weighted by abundances, i.e.:
\begin{equation}\label{eq.LMM}
  \mathbf{x} = \mathbf{Ma}+\mathbf{n}
\end{equation}
where $\mathbf{x}\in\mathbb{R}^{B\times 1}$ is a vector containing $B$ spectral bands associated with the obtained pixel, $\mathbf{M}\in\mathbb{R}^{B\times R}$ is the endmember matrix with $R$ spectral signatures of pure constituents, $\mathbf{a}\in\mathbb{R}^{R\times 1}$ is the corresponding abundance, $\mathbf{n}$ denotes the additive noise. Due to the physical interpretation of hyperspectal image and mixing model, the endmembers and abundances are considered to satisfy the nonnegativity constraint (ENC and ANC), and the abundances are also assumed to satisfy sum-to-one for each pixel (ASC). It is important to highlight that the ASC can be ignored for some physically motivated reasons. This may occur, for instance, when there are local variations in the topography of the scene~\cite{drumetz2016blind}. In our work, we take into account all these constraints, i.e.,
  \begin{align}\label{eq.AEC}
    \mathcal{D}_{M} & :=\{\mathbf{M}|\mathbf{M}\succeq\mathbf{0}\} \\
    \mathcal{D}_{A} & :=\left\{\mathbf{a}|\mathbf{a}\succeq\mathbf{0}, \sum_{i=1}^{R}a_{i}=1\right\}
  \end{align}
in which $\mathcal{D}_{M}$ and $\mathcal{D}_{A}$ are the feasible regions of endmembers and abundances, respectively.

For a hyperspectral image $\mathbf{X}\in\mathbb{R}^{B\times N}$ with $N$ pixels, the hyperspectral unmixing task is formulated by the following optimization problem:
\begin{equation}\label{eq.unmix}
  \min_{\substack{\mathbf{M}\in\mathcal{D}_{M}\\ \mathbf{A}\in \mathcal{D}_{A}}}
  \frac{1}{2}\|\mathbf{X}-\mathbf{MA}\|_{\text{F}}^2+\lambda\mathcal{R}(\mathbf{M},\mathbf{A})
\end{equation}
where the first term is the data fitting term, the second term serves as regularization aiming to enforce certain desirable properties of endmembers and abundances, and $\lambda$ is the trade-off parameter. Different regularization techniques have been devised.
Conventional methods typically design regularization according to the spatial and spectral characteristics in hyperspectral images. For instance, in most scenes, the pixel values of an image are piece wise continuous in the spatial dimension. The works~\cite{iordache2012total,xiong2018hyperspectral,li2020superpixel} introduce total variation (TV) to achieve this purpose. As a pixel usually contains much fewer materials than the number of pure materials contained in the endmember matrix, sparse constraints are introduced to the abundance matrix to obtain sparse results~\cite{iordache2012total,li2020superpixel}. Minimum-volume simplex regularization is incorporated in the objective function to constrain the volume of the simplex formed by endmembers~\cite{li2015minimum}.

Typically, the alternative direction method of multipliers (ADMM) is a powerful tool to solve problem~\eqref{eq.unmix}, which decomposes a complicated optimization problem into several easier subproblems. The work~\cite{iordache2012total} utilises this strategy to address the objective function with two regularizations on abundance, namely $\ell_1$-norm and TV. A nonstandard application of ADMM with a block coordinate descent
scheme is designed to address a 3D TV constrained problem, which can model the spatial and spectral correlations and gain sharper edges~\cite{wang2019blind}.
In~\cite{qin2020blind}, an ADMM based blind hyperspetral unmixing method is proposed, which can simultaneously estimate abundances and endmembers. A graph TV regularization is considered to capture the spatial correlation information. The work~\cite{li2019graph} uses this tool to solve a graph regularized nonlinear unmixing method with the multilinear mixing model to indicate the nonlinearity between endmembers. Many other ADMM-based unmixing methods have also been proposed, including those for nonlinear unmixing using kernel methods~\cite{gu2021nonlinear} and those addressing unmixing problems that consider spectral variability~\cite{thouvenin2015hyperspectral}.
While ADMM is a versatile and powerful tool for solving the unmixing optimization problems, the choice of penalty parameters inherently limits the performance.

Compared to traditional methods that rely on manually designed regularization terms, deep neural networks (DNNs) can automatically extract and model complex patterns and nonlinear relationships in data. DNNs have achieved breakthrough results in hyperspectral image processing tasks such as object recognition~\cite{yang2022deep}, band selection~\cite{cai2019bs}, and image super-resolution~\cite{cheng2024general}.
The encoding and decoding processes of autoencoder perfectly fits the formulation of the unmixing problems, and this kind of methods have achieved great advances~\cite{chen2023integration}. Thus many unmixing methods have been proposed  based on deep autoencoders~\cite{palsson2022blind}.
To effectively leverage the spatial structure information of hyperspectral image, deep convolutional neural network (CNN) based autoencoders are conducted~\cite{palsson2020convolutional,khajehrayeni2020hyperspectral}. Some unmixing networks with modified autoencoder architectures have also been proposed. For instance, MSSS-Net~\cite{qi2023multiview} builds a two-stream unmixing framework, which adopts an end-to-end manner to simultaneously learn spatial stream and multi-view spectral stream networks, and fuses information of different scales to achieve more effective unmixing.
In work~\cite{hong2021endmember}, an endmember-guided subnetwork is introduced alongside the basic autoencoder framework to extract features from pure or near-pure spectra. Meanwhile, a weight sharing strategy is used to guide the learning of the basic network. However, DNNs are often regarded as ``black boxes" and lack of physical interpretability.

Recently, integration model- and deep learning- based methods is a new trend to design unmixing frameworks with both interpretability and data driven advantages. On one hand, the plug-and-play unmixing approaches use a pretrained denoiser to provide priors~\cite{wang2020hyperspectral,wang2023tuning}. However, these methods still require setting the value of penalty parameters. On the other hand, unrolling/unfolding the optimization algorithm to design the unmixing network structure receives great performance~\cite{cui2023unrolling,shao2023iviu,zhou2021admm,kong2023deep}, which breaks through the drawbacks of iterative algorithms and plugs interpretability to the deep architecture. The work~\cite{zhou2021admm} unrolls the constrained sparse regrassion (CSR) problem to construct the abundance estimation network, which has shown better performance with lighter structures and faster convergence speed. The work~\cite{zhou2021admm} employs a fully CNN to unroll the variable splitting and augmented Lagrangian algorithm. However, the spatial information learned by this model is still unsatisfactory, and its performance is limited in high-noise scenes.

In our work, to overcome the poor interpretability of deep learning based unmixing architecture, we propose a novel unmixing network, named PnP-Net. Our method unrolls the plug-and-play framework and uses pretrained state-of-the-art denoisers to add information learnt from large scale of image datasets.
The main merits of this work are threefold:
\begin{itemize}
  \item Our proposed method can take advantages of both the optimization- and learning-based methods. We unroll the plug-and-play unmixing framework into a novel neural network scheme, which can be trained in an end-to-end manner. Our approach relies on the ADMM algorithm, and the layers mimic the iterative process.
  \item By using the denoiser pretrained with external datasets, our proposed framework effectively leverages the information from additional data and combines the priors with internal learning from the hyperspectral data. This strategy bypasses the difficulty of limited hyperspectral data and the training of a large number of network parameters.
  \item In this work, we design a dynamic convolution using kernels of various sizes to capture multiscale features for spectral unmixing. It increases the model complexity with multiple parallel convolution kernels fused by an attention module.
\end{itemize}

The remainder of this paper is organized as follows. We describe the background and related works of our method in Section~\ref{sec:related_work}. In Section~\ref{sec:proposed}, we present our proposed PnP-Net method in detail and give the main flowchart and training details. Section~\ref{sec:exp} shows the experiment results to validate the effectiveness of our method. Section~\ref{sec:conclusion} concludes this work.
\section{Related Works}
\label{sec:related_work}
In this section, we briefly review the basic concepts of the plug-and-play unmixing framework and deep unrolling network.
\subsection{Plug-and-Play Unmixing Methods}
The plug-and-play framework benefits from variable splitting technique, allowing the utilization of denoising priors to effectively tackle a range of image restoration problems. It has been used for hyperspectral unmixing~\cite{zhao2021plug}. Typically, the denoising regularization is plugged for the abundances with \eqref{eq.unmix} rewritten as:
\begin{equation}\label{eq.unmix_1}
  \min_{\substack{\mathbf{M}\in\mathcal{D}_{M}\\ \mathbf{A}\in \mathcal{D}_{A}}}
  \frac{1}{2}\|\mathbf{X}-\mathbf{MA}\|_{\text{F}}^2+\lambda\mathcal{R}(\mathbf{A}).
\end{equation}
We use ADMM to resolve the optimization problem described in \eqref{eq.unmix_1}. By introducing an auxiliary variable $\mathbf{V}$ to replace $\mathbf{A}$ in the regularization term of \eqref{eq.unmix_1} and adding constraint $\mathbf{V}=\mathbf{A}$, the original problem is decomposed into easier and more manageable subproblems. The iterative optimization process of \eqref{eq.unmix_1} is as follows:
\begin{align}\label{eq.pnp}
  \left\{\mathbf{M},\mathbf{A}\right\}&\leftarrow\arg\min_{\substack{\mathbf{M}\in\mathcal{D}_{M}\\ \mathbf{A}\in \mathcal{D}_{A}}} \|\mathbf{X}-\mathbf{MA}\|_{\text{F}}^2+\frac{\rho}{2}\|\mathbf{A}-\mathbf{V}+\mathbf{G}\|_{\text{F}}^2\\
  \mathbf{V}& \leftarrow\arg\min_{\mathbf{V}}\frac{\rho}{2}\|\mathbf{A}-\mathbf{V}+\mathbf{G}\|_{\text{F}}^2+\lambda\mathcal{R}(\mathbf{V})\\
  \mathbf{G}&\leftarrow\mathbf{G}+\mathbf{A}-\mathbf{V}
\end{align}
in which $\mathbf{G}$ is the dual variable, and $\rho$ is the penalty parameter. The process involves two essential operators. The first is a blind unmixing operator to estimate endmembers and abundances. The second step can be viewed as a denoising of $\mathbf{A}+\mathbf{G}$. The regularization term $\mathcal{R}(\mathbf{V})$ can be implicitly coped by incorporating a denoising operator ($\mathsf{C}(\cdot)$), i.e.,
\begin{equation}\label{eq.pnp_d}
\mathbf{V}=\mathsf{C}\left(\mathbf{A}+\mathbf{G}\right).
\end{equation}
In general, the denoiser $\mathsf{C}(\cdot)$ can be any readily available denoising operator. For example, the well-known nonlocal means denoising (NLM)~\cite{buades2011non} and block-matching and 3D filtering (BM3D)~\cite{dabov2007image} have been plugged to capture image priors and get high-quality unmixing results. This provides the possibility of integrating CNN-based denoisers with robust prior learning from mass image data and addressing the drawback of insufficient volume of hyperspectral data.
\subsection{Deep Unrolling Network}
The deep unrolling paradigm involves unfolding iterative optimization algorithms into trainable deep architectures. One popular deep unrolling unmixing method is unfolding the CSR for sparse unmixing~\cite{zhou2021admm}. With known $\mathbf{M}$ and $\mathcal{R}(\mathbf{A})=\|\mathbf{A}\|_1$, it introduces an
auxiliary variable $\mathbf{V}=\mathbf{A}$. The ADMM iteratively addresses the optimization problem through the following steps:
\begin{align}
\mathbf{A}^{(k+1)}&=\mathbf{W}\mathbf{X}+\mathbf{Q}\left(\mathbf{V}^{(k)}-\mathbf{G}^{(k)}\right)\label{eq.unroll_1} \\
  \mathbf{V}^{(k+1)}&=\max\left(\text{soft}\left(\mathbf{A}^{(k+1)}+\mathbf{G}^{(k)},\frac{\lambda}{\rho}\right),\mathbf{0}\right)\label{eq.unroll_2}\\
  \mathbf{G}^{(k+1)}&=\mathbf{G}^{(k)}+\mathbf{A}^{(k+1)}-\mathbf{V}^{(k+1)} \label{eq.unroll_3}
\end{align}
in which $\mathbf{W}=\left(\mathbf{M}^{\top}\mathbf{M}+\rho\mathbf{I}\right)^{-1}\mathbf{M}^{\top}$ and $\mathbf{Q}=\left(\mathbf{M}^{\top}\mathbf{M}+\rho\mathbf{I}\right)^{-1}\rho$. $\text{Soft}(a,\theta)=\text{sign}(a)(|a|-\theta)_{+}$ is the soft-threshold operator to resolve the $\ell_1$-norm. The iterative steps from \eqref{eq.unroll_1} to \eqref{eq.unroll_3} can be unfolded into network layers, denoted as $\text{Layer}_A$, $\text{Layer}_V$ and $\text{Layer}_G$ with learnable parameters $\left\{\mathbf{W}^{(k+1)},\mathbf{Q}^{(k+1)},\theta^{(k+1)}\right\}$, where $\theta^{(k+1)}$ replaces the role of $\frac{\lambda}{\rho}$. Thus \eqref{eq.unroll_1} to \eqref{eq.unroll_3} can be reexpressed as follows:
\begin{equation}
  \begin{split}
 \mathbf{A}^{(k+1)} & = \text{Layer}_{A}\left(\mathbf{V}^{(k)},\mathbf{G}^{(k)};\mathbf{W}^{(k+1)},\mathbf{Q}^{(k+1)}\right) \\
  &=\mathbf{W}^{(k+1)}\mathbf{X}+\mathbf{Q}^{(k+1)}\left(\mathbf{V}^{(k)}-\mathbf{G}^{(k)}\right)\\
\end{split}
\end{equation}
\begin{equation}
\begin{split}
  \mathbf{V}^{(k+1)} & =\text{Layer}_{V}\left(\mathbf{A}^{(k+1)},\mathbf{V}^{(k)},\theta^{(k+1)}\right) \\
    &= \text{ReLU}\left(\mathbf{A}^{(k+1)}+\mathbf{G}^{(k)}-\theta^{(k+1)}\mathbf{I}\right)
\end{split}
\end{equation}
\begin{equation}
\begin{split}
  \mathbf{G}^{(k+1)} & =\text{Layer}_G\left(\mathbf{A}^{(k+1)},\mathbf{V}^{(k+1)},\mathbf{G}^{(k)}\right) \\
    &= \mathbf{G}^{(k)}+\mathbf{A}^{(k+1)}-\mathbf{V}^{(k+1)}.
\end{split}
\end{equation}
Each iteration is decomposed into a single network layer, and combining these layers many times is analogous to performing multiple iterations of CSR. The unrolling deep network can be trained by an end-to-end manner. By this way, the knowledge of classical iterative algorithms are infused in the design of deep network for hyperspectral unmixing. However, the networks obtained by unrolling iterative algorithms are restricted to one pixel only and not take into account of spatial structure information.
By some modifications of networks, we can employ the well-designed layers such as convolutional networks to add more spatial priors.

\section{Proposed Method}
\label{sec:proposed}
\subsection{Problem Formulation}
In \eqref{eq.unmix}, handcrafting an effective regularizer $\mathcal{R}(\mathbf{M},\mathbf{A})$ and devising an efficient algorithm to solve the objective function are challenging tasks. Rather than using this intricate path, we propose to derive priors from various image data and integrate them into the optimization process by a plug-and-play strategy, where the regularization is implicitly designed by the utilization of a denoising algorithm. In hyperspectral unmixing task, the spatial information is embedded in the abundance, and the spectral information are contained in the endmember. Thus we plug regularization on abundance to fully exploit the spatial information.
We introduce regularization by denoising (RED) as the regularizer to add denoising priors, and the corresponding hyperspectral unmixing optimization problem can be reformulated as:
\begin{equation}\label{eq.pro_1}
  \min_{\mathbf{M},\mathbf{A}}
  \frac{1}{2}\|\mathbf{X}-\mathbf{MA}\|_{\text{F}}^2+\lambda\mathcal{R}_{\text{RED}}(\mathbf{A})
  +\Omega_{\mathcal{D}_{A}}(\mathbf{A})+\Omega_{\mathcal{D}_{M}}(\mathbf{M}),
\end{equation}
where $\Omega_{\mathcal{D}}$ is an indicator function of a set $\mathcal{D}$ and is defined as:
\begin{equation}
\Omega_{\mathcal{D}}(\mathbf{Z})= \begin{cases}0, & \mathbf{Z} \in \mathcal{D} \\ \infty, & \text { otherwise. }\end{cases}
\end{equation}
$\mathcal{R}_{\text{RED}}(\mathbf{A})$ is defined as:
\begin{equation}\label{eq.pro_2}
  \mathcal{R}_{\text{RED}}(\mathbf{A})=\frac{1}{2}\mathbf{A}^{\top}\left(\mathbf{A}-\mathsf{C}(\mathbf{A})\right),
\end{equation}
where $\mathsf{C}(\cdot)$ is an off-the-shelf denoiser. RED depends on the inner product between the solution $\mathbf{A}$ and its residual after denoising $\mathbf{A}-\mathsf{C}(\mathbf{A})$. Under mild assumptions, it demonstrates advantageous derivative properties with $\nabla \mathcal{R}(\mathbf{A})=\mathbf{A}-\mathsf{C}(\mathbf{A})$. This formulation also provides flexibility in accommodating any denoising engines like original plug-and-play denoising framework.

We use ADMM algorithm to solve \eqref{eq.pro_1}. Two auxiliary variables $\mathbf{V}_1$ and $\mathbf{V}_2$ are introduced, and the objective function is rewritten as:
\begin{equation}\label{eq.pro_3}
\begin{split}
  \min_{\substack{\mathbf{M},\mathbf{A},\\\mathbf{V}_1,\mathbf{V}_2}}
  \frac{1}{2}\|\mathbf{X}-&\mathbf{M}\mathbf{A}\|_{\text{F}}^2+\lambda\mathcal{R}_{\text{RED}}(\mathbf{V}_1)
  +\Omega_{\mathcal{D}_{A}}(\mathbf{A})
  +\Omega_{\mathcal{D}_{M}}(\mathbf{V}_2)\\
    \text{s.t.}&~~\mathbf{A}=\mathbf{V}_1,~\mathbf{M}=\mathbf{V}_2.
\end{split}
\end{equation}
The associated augmented Lagrangian is expressed as:
\begin{equation}\label{eq.pro_4}
\begin{split}
  \mathcal{J}=\frac{1}{2}\|\mathbf{X}-&\mathbf{M}\mathbf{A}\|_{\text{F}}^2+\lambda\mathcal{R}_{\text{RED}}(\mathbf{V}_1)+\Omega_{\mathcal{D}_{A}}(\mathbf{A})
  +\Omega_{\mathcal{D}_{M}}(\mathbf{V}_2)\\
  &+\frac{\alpha}{2}\|\mathbf{A}-\mathbf{V}_1+\mathbf{G}_1\|_{\text{F}}^2+
  \frac{\beta}{2}\|\mathbf{M}-\mathbf{V}_2+\mathbf{G}_2\|_{\text{F}}^2
\end{split}
\end{equation}
where $\mathbf{G}_1$ and $\mathbf{G}_2$ are dual variables, and $\alpha$ and $\beta$ are nonnegative parameters. Then the ADMM algorithm involves solving four subproblems at each iteration. In our work, we unroll the ADMM algorithm into a deep network by associating each iteration with a single network layer and then stacking a finite number of layers together. Through unrolling, we can use back-propagation to learn regularization parameters that are difficult to handcraft. This approach allows us to capture image priors through a well-designed network structure and add physical interpretability to the deep network. The proposed unrolling layers (PnP-Net) are presented in detail as follows.
\subsection{Unrolling Plug-and-play Framework for A Layer}
\subsubsection{Update $\mathbf{A}$}
\begin{figure}[t]
  \centering
  \includegraphics[width=8cm]{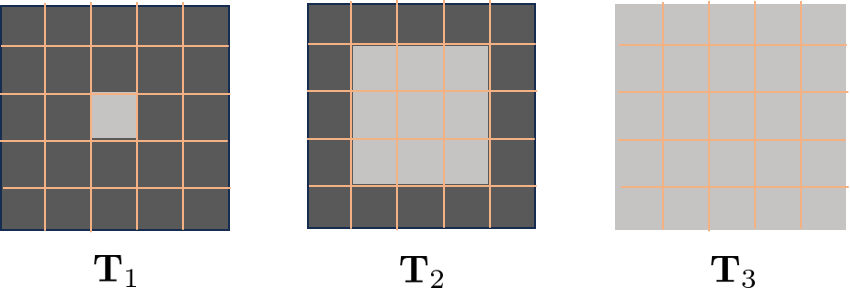}\\
  \caption{A diagram of the $\mathbf{T}$ matrix. $L=3$ for example with kernel sizes set to $1\times 1$,  $3\times 3$ and $5\times 5$. The black blocks represent the corresponding weights are zeros, the gray blocks are with available weights, and $\{\mathbf{T}_l\}_{l=1}^{L}(i,j)=1$.}\label{fig.T_figure}
\end{figure}

\begin{figure}[t]
  \centering
  \includegraphics[width=8cm]{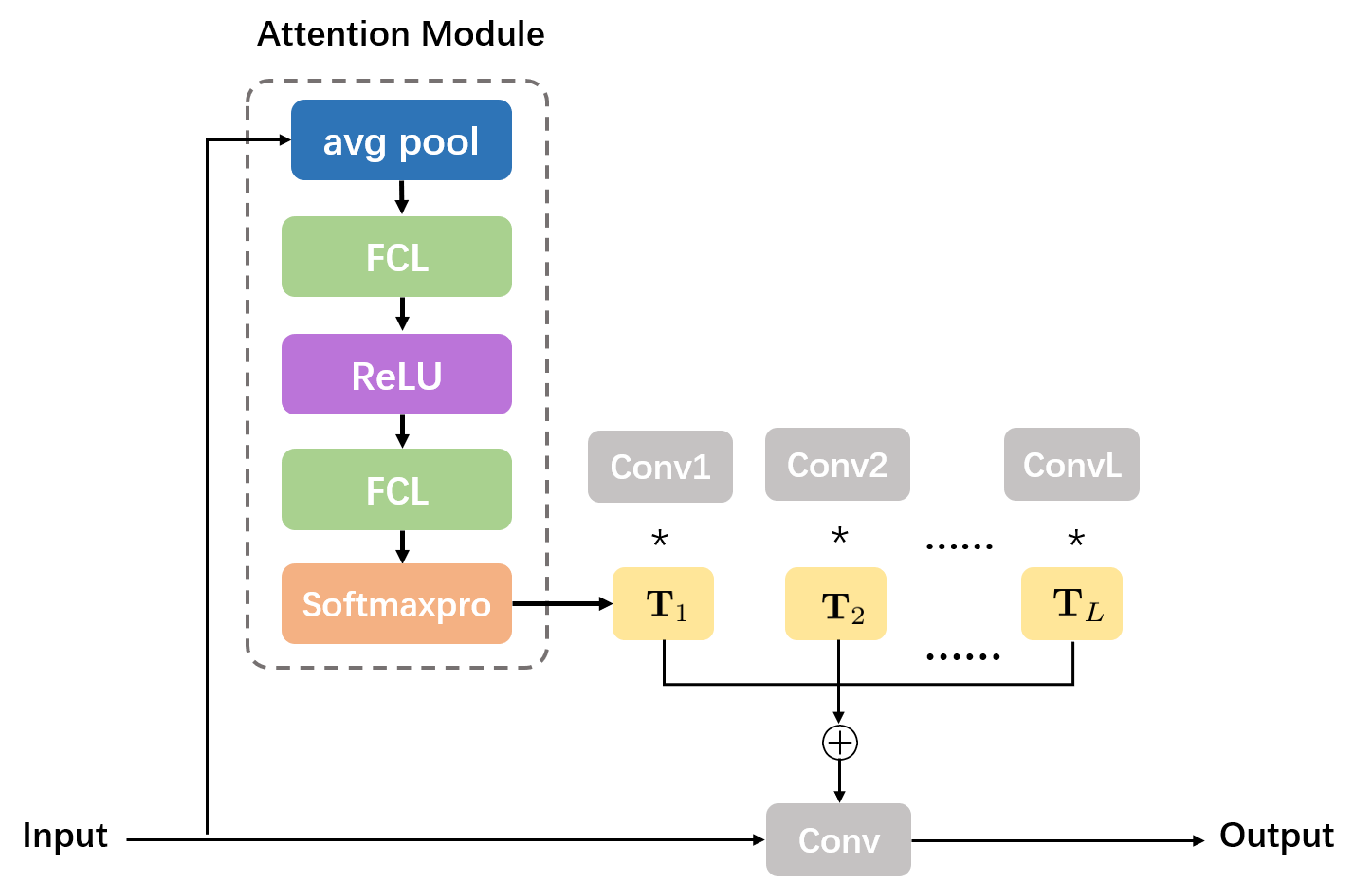}\\
  \caption{A scheme of the dynamic convolution layer.}\label{fig.cnn_layer}
\end{figure}

The $\Omega_{\mathcal{D}_{A}}$ can be satisfied with a Softmax operator. The remained part of $A$-subproblem is a least-square problem defined as
\begin{equation}\label{eq.pro_5_A}
\min_{\mathbf{A}}\frac{1}{2}\|\mathbf{X}-\mathbf{M}\mathbf{A}\|_{\text{F}}^2+\frac{\alpha}{2}\|\mathbf{A}-\mathbf{V}_1+\mathbf{G}_1\|_{\text{F}}^2.
\end{equation}
It can be solved with a closed-form as follws:
\begin{equation}\label{eq.pro_5_A_2}
\mathbf{A}^{(k+1)}=\left(\mathbf{M}^{\top}\mathbf{M}+\alpha\mathbf{I}\right)^{-1}
\left[\mathbf{M}^{\top}\mathbf{X}+\alpha\left(\mathbf{V}_1-\mathbf{G}_1\right)\right].
\end{equation}
The $A$-update layer is designed by unfolding \eqref{eq.pro_5_A_2} and rewrites as
\begin{equation}\label{eq.pro_5_A_3}
\begin{split}
  \mathbf{A}^{(k+1)}&=\text{Layer}_A\left(\mathbf{V}_1^{(k)},\mathbf{G}_1^{(k)},\mathbf{X},
  \mathbf{W}_1,\mathbf{Q}_1\right)\\
  &=\mathbf{W}_1\mathbf{X}+\mathbf{Q}_1\left(\mathbf{V}_{1}^{(k)}-\mathbf{G}_{1}^{(k)}\right)
\end{split}
\end{equation}
where
\begin{equation}\label{eq.pro_5_A_3_2}
\begin{split}
\mathbf{W}_1&=\left(\mathbf{M}^{\top}\mathbf{M}+\alpha\mathbf{I}\right)^{-1}\mathbf{M}^{\top}\\
\mathbf{Q}_1&=\left(\mathbf{M}^{\top}\mathbf{M}+\alpha\mathbf{I}\right)^{-1}\alpha.
\end{split}
\end{equation}
As for a closed-form solution, conventional unrolling methods only learn the regularization parameters. In~\cite{zhou2021admm} and \cite{kong2023deep}, $\mathbf{W}_1$ and $\mathbf{Q}_1$ are considered as the learnable parameters to enhance the flexibility of network and improve the performance. Following this strategy, we also replace the fixed $\mathbf{W}_1$ and $\mathbf{Q}_1$ as parameters to estimate.

In~\cite{zhou2021admm}, two fully connected layers with bias set to $\mathbf{0}$ adding together are used to formulate $\text{Layer}_A$. In~\cite{kong2023deep}, 2D convolutional layers are used to generate this layer, which can capture spatial information. In our work, inspired by the dynamic convolution work~\cite{chen2020dynamic}, we design a novel dynamic convolution with multiscale convolution kernels to formulate this layer, which can capture abundant spatial information from different scales. This layer consolidates multiple parallel convolution kernels dynamically, adjusting their contributions based on input-dependent attentions. Instead of enhancing either the depth or the width of the network, this layer possesses increased representational capability due to the nonlinear aggregation of these kernels through attention. Convolutional kernels with different sizes are able to capture multiscale information. We denote the traditional 2D convolution layer as:
\begin{equation}
\mathbf{O}=\mathbf{K}\otimes\mathbf{Ix}
\end{equation}
where $\mathbf{K}$ is the convolutional kernel, $\mathbf{Ix}$ denotes the input, and $\mathbf{O}$ is the output. Our dynamic convolution layer (DCL) with $L$ parallel multiscale convolution kernels is defined as:
\begin{equation}
\mathbf{O}=\sum_{l=1}^{L}\left(\mathcal{T}(\mathbf{T}_l)\odot\mathbf{K}_l\right)\otimes\mathbf{Ix}
\end{equation}
where $\mathbf{T}_l$ is the attention weight of the $l{\text{th}}$ convolution layer, the size of $\mathbf{T}_l$ is the same as the maximum size of convolutional kernels. The size of available weights of $\mathbf{T}_l$ is the same as the $l{\text{th}}$ kernel's, and they are designed on the centre of $\mathbf{T}_l$. Other positions are padded with zeros.
The summation of each position of $\mathbf{T}$ is one, i.e. $\{\mathbf{T}_l\}_{l=1}^{L}(i,j)=1$ (the $(i,j)$th position).
The size of $\mathbf{T}_l$ is inconsistent
with $\mathbf{K}_l$.
We use $\mathcal{T}$ to extract the available weights from $\mathbf{T}_l$ to make $\mathbf{T}_l$ and $\mathbf{K}_l$ the same size.
$\odot$ is an element-by-element multiplication. An illustration of $\mathbf{T}$ matrix is shown in Fig.~\ref{fig.T_figure}.
Then the output $\mathbf{O}$ represents the optimal combination of multiscale features. The weight $\mathbf{T}$ is associated with the input and learnt by an attention module, and the squeeze-and-excitation module~\cite{hu2018squeeze} is introduced to calculate it. The global average pooling is firstly applied to squeeze the global spatial information of an input data cube. Then two fully connected layers with an ReLU between them as the activated function are exploited to extract features. A specially designed layer, named Softmaxpro, is used to obtain sum-to-one attention weights $\mathbf{T}$ with the characteristics described above. 
To be specific,
for the $\mathbf{T}_l$, a fully connected layer is used to map the size of features the same as the $l$th convolutional kernel. Then a zero-padding operator makes it the same as the maximum size of convolutional kernels. Finally, we apply the Softmax to the available weights of each position of $\mathbf{T}$ to accomplish this goal:
\begin{equation}\label{eq.softmax}
  \{\mathbf{T}_l\}_{l=1}^{L}(i,j)=\text{Softmax}\left(\left\{\mathcal{T}(\mathbf{T}_l)\right\}_{l=1}^{L}(i,j)\right).
\end{equation}
The scheme of our proposed DCL is shown in Fig.~\ref{fig.cnn_layer}.

With this useful and efficient network component, \eqref{eq.pro_5_A_3} is equivalently reformulated as:
\begin{equation}\label{eq.pro_5_A_3_1}
\begin{split}
\mathbf{A}^{(k+1)}=&\text{Layer}_A\left(\mathbf{V}_1^{(k)},\mathbf{G}_1^{(k)},\mathbf{X};
  \left\{\mathbf{W}_1,\mathbf{Q}_1, \mathbf{R}, \mathbf{H}\right\}^{(k+1)}\right)\\
  =&\sum_{l=1}^{L}\left(\mathcal{T}\left(\mathbf{R}_{l}^{(k+1)}\right)\odot\mathbf{W}_{1,l}^{(k+1)}\right)\otimes\mathbf{X}\\
  &+\sum_{p=1}^{P}\left(\mathcal{T}\left(\mathbf{H}_{p}^{(k+1)}\right)\odot\mathbf{Q}_{1,p}^{(k+1)}\right)\otimes\left(\mathbf{V}_{1}^{(k)}-\mathbf{G}_{1}^{(k)}\right),
\end{split}
\end{equation}
where $L$ and $P$ are the number of kernels, $\mathbf{W}_{1,l}$ and $\mathbf{Q}_{1,l}$ represent the weights of the $l$th convolution kernel, and $\mathbf{R}_l$ and $\mathbf{H}_p$ are the attention weight matrices of dynamic convolution. 

\begin{figure}
  \centering
  \includegraphics[width=9cm]{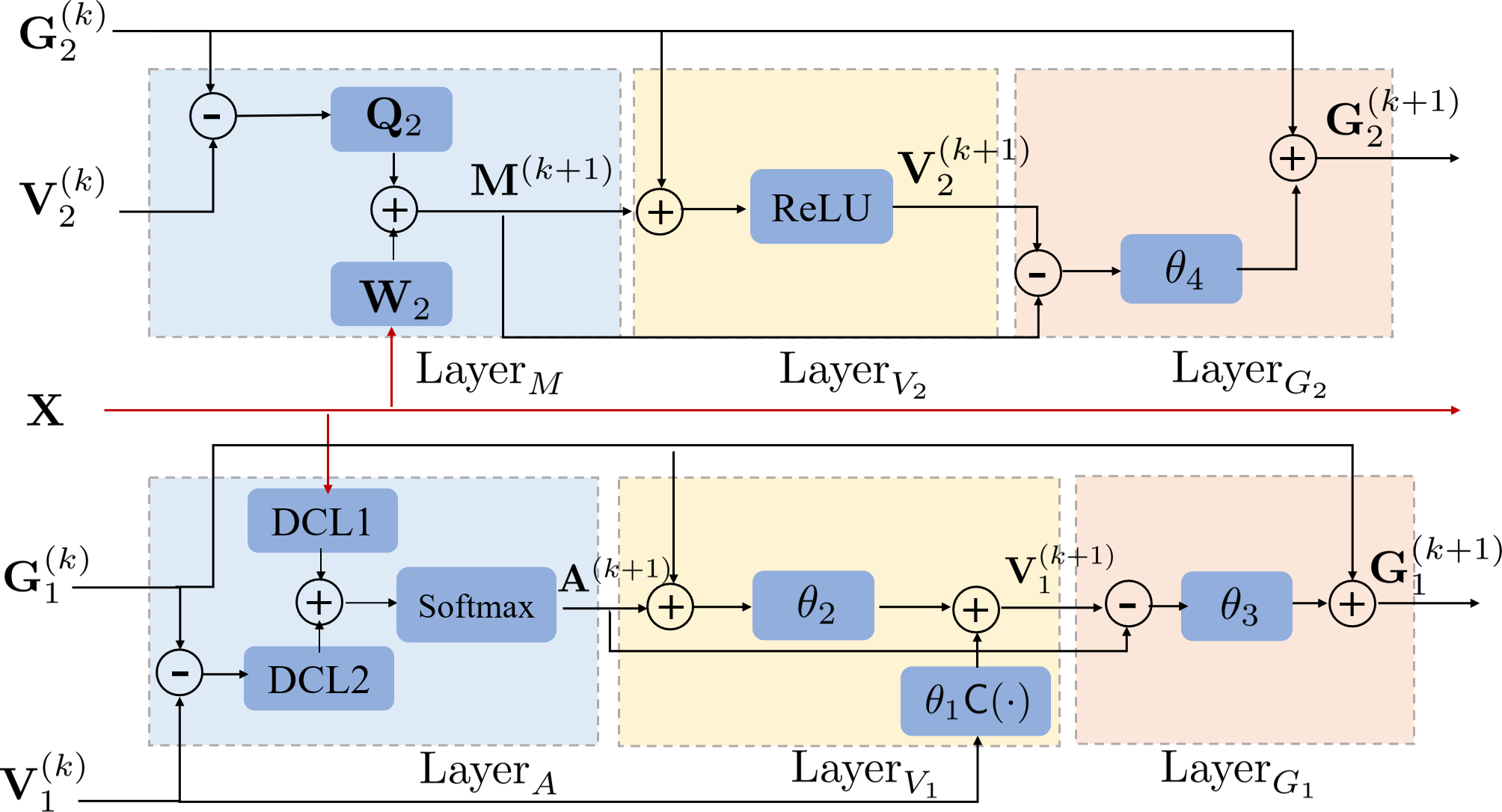}\\
  \caption{The architecture of an unrolling layer (one block).}\label{fig.unroll}
\end{figure}

Then we apply a Softmax operator to constrain the output of $\text{Layer}_A$ within $\mathcal{D}_A$, i.e.,
\begin{equation}\label{eq.pro_5_A_4}
  \mathbf{A}=\text{Softmax}\left(\text{Layer}_A\left(\mathbf{V}_1^{(k)},\mathbf{G}_1^{(k)},\mathbf{X};
  \left\{\mathbf{W}_1,\mathbf{Q}_1\right\}^{(k+1)}\right)\right).
\end{equation}
The $\text{Layer}_{A}$ neural network structure is illustrated in Fig.~\ref{fig.unroll}.
\subsubsection{Update $\mathbf{V}_1$} The $V_1$-subproblem is a standard RED objective function:
\begin{equation}\label{eq.pro_6_V1}
\min_{\mathbf{V}_1}\frac{\alpha}{2}\|\mathbf{A}-\mathbf{V}_1+\mathbf{G}_1\|_{\text{F}}^2
+\lambda\mathcal{R}_{\text{RED}}(\mathbf{V}_1).
\end{equation}
It also can be solved in a closed-form.
We use the fixed-point strategy to solve this problem. By setting the
gradient of the objective function to $\mathbf{0}$, we have the following equation:
\begin{equation}\label{eq.pro_6_V1_2}
\lambda\left(\mathbf{V}_1-\mathsf{C}(\mathbf{V}_1)\right)+\alpha\left(\mathbf{A}-\mathbf{V}_1+\mathbf{G}_1\right)=\mathbf{0}.
\end{equation}
The solution of $V_1$-subproblem is an iterative scheme as follows:
\begin{equation}\label{eq.pro_6_V1_3}
\mathbf{V}_1^{(k+1)}=\frac{1}{\lambda+\alpha}\left[\lambda\mathsf{C}(\mathbf{V}_1^{(k)})+\alpha(\mathbf{A}^{(k+1)}+\mathbf{G}_1^{(k)})\right].
\end{equation}
The $V_1$-update layer is derived by unrolling \eqref{eq.pro_6_V1_3}. The update of the $(k+1)$th iteration of $\text{Layer}_{V_1}$ is expressed as
\begin{equation}\label{eq.pro_6_V1_4}
\begin{split}
\mathbf{V}_1^{(k+1)}&=\text{Layer}_{V_1}\left(\mathbf{A}_1^{(k+1)},\mathbf{G}_1^{(k)};\theta_1^{(k+1)},\theta_2^{(k+1)}\right)\\
&=\theta_1^{(k+1)}\mathsf{C}(\mathbf{V}_1^{(k)})+\theta_2^{(k+1)}(\mathbf{A}^{(k+1)}+\mathbf{G}_1^{(k)})
\end{split}
\end{equation}
where $\theta_1^{(k+1)}$ and $\theta_2^{(k+1)}$ are learnable parameters and play the role of $\frac{\lambda}{\lambda+\alpha}$ and $\frac{\alpha}{\lambda+\alpha}$. In this manner, we can use a data-driven strategy to learn the regularization parameters, rather than handcraft them.
The $\text{Layer}_{V_1}$ neural network layer is shown in Fig.~\ref{fig.unroll}.
\subsubsection{Update $\mathbf{M}$}
As for the update of $M$-subproblem, it aims to solve the following optimization problem:
\begin{equation}\label{eq.pro_7_M_1}
  \min_{\mathbf{M}}\frac{1}{2}\|\mathbf{X}-\mathbf{M}\mathbf{A}\|_{\text{F}}^2
  +\frac{\beta}{2}\|\mathbf{M}-\mathbf{V}_2+\mathbf{G}_2\|_{\text{F}}^2,
\end{equation}
which can be efficiently solved through a closed-form solution. Then we obtain the following update iteration:
\begin{equation}\label{eq.pro_7_M_2}
  \mathbf{M}^{(k+1)}=\left(\mathbf{X}\mathbf{A}^{\top}-\beta\mathbf{V}_2+\beta\mathbf{G}_2\right)\left(\mathbf{A}\mathbf{A}^{\top}-
  \beta\mathbf{I}\right)^{-1}.
\end{equation}

The $M$-update layer is designed by unfolding \eqref{eq.pro_7_M_2}. With the $\mathbf{V}_2^{(k)}$, $\mathbf{G}_2^{(k)}$ and given $\mathbf{X}$, it is written as follows:
\begin{equation}\label{eq.pro_7_M_3}
\begin{split}
\mathbf{M}^{(k+1)}&=\text{Layer}_{M}\left(\mathbf{V}_2^{(k)},\mathbf{G}_2^{(k)},\mathbf{X};
  \mathbf{W}_2,\mathbf{Q}_2\right)\\
&=\mathbf{X}\mathbf{W}_2+\left(\mathbf{G}_2^{(k)}-\mathbf{V}_2^{(k)}\right)\mathbf{Q}_2
\end{split}
\end{equation}
where
\begin{equation}\label{eq.pro_7_M_4}
\begin{split}
\mathbf{W}_2&=\mathbf{A}^{\top}\left(\mathbf{A}\mathbf{A}^{\top}-\beta\mathbf{I}\right)^{-1}\\
\mathbf{Q}_2&=\left(\mathbf{A}\mathbf{A}^{\top}-
  \beta\mathbf{I}\right)^{-1}\beta.
\end{split}
\end{equation}
We replace $\mathbf{W}_2$ and $\mathbf{Q}_2$ with learnable parameters to enhance flexibility. The $\text{Layer}_{M}$ neural network layer is presented in Fig.~\ref{fig.unroll} and defined as
\begin{equation}\label{eq.pro_7_M_3_1}
\mathbf{M}^{(k+1)}=\text{Layer}_{M}\left(\mathbf{V}_2^{(k)},\mathbf{G}_2^{(k)},\mathbf{X};
  \left\{\mathbf{W}_2,\mathbf{Q}_2\right\}^{(k+1)}\right).
\end{equation}

\begin{figure*}[t]
  \centering
  \includegraphics[width=17cm]{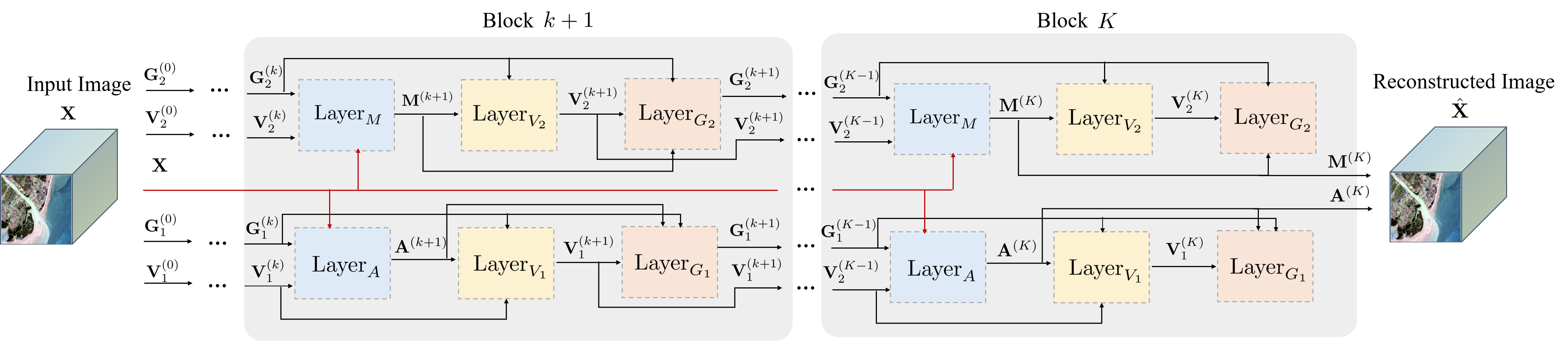}\\
  \caption{Architecture of the proposed PnP-Net with $K$ blocks.}\label{fig.framework}
\end{figure*}
\subsubsection{Update $\mathbf{V}_2$}
The $V_2$-subproblem is a least-square problem with the nonnegative constraint, which is formulated as:
\begin{equation}\label{eq.pro_8_V2_1}
\min_{\mathbf{V}_2}\frac{\beta}{2}\|\mathbf{M}-\mathbf{V}_2+\mathbf{G}_2\|_{\text{F}}^2
  +\Omega_{\mathcal{D}_{M}}(\mathbf{V}_2).
\end{equation}
We use a hard threshold operator to solve this problem by mapping $\mathbf{M}+\mathbf{G}_2$ to a set of all nonnegative elements. This operation is written as:
\begin{equation}\label{eq.pro_8_V2_2}
\mathbf{V}_2 = \max\left(\mathbf{M}+\mathbf{G}_2,\mathbf{0}\right).
\end{equation}

In the design of $\text{Layer}_{V_2}$, we use the ReLU to generate this layer:
\begin{equation}\label{eq.pro_8_V2_3}
\begin{split}
\mathbf{V}_2^{(k+1)}&=\text{Layer}_{V_2}\left(\mathbf{M}^{(k+1)},\mathbf{G}_2^{(k)}\right)\\
&=\text{ReLU}\left(\mathbf{M}^{(k+1)}+\mathbf{G}_2^{(k)}\right).
\end{split}
\end{equation}
The $\text{Layer}_{V_2}$ neural network layer is shown in Fig.~\ref{fig.unroll}.
\subsubsection{Update $\mathbf{G}_1$ and $\mathbf{G}_2$}
The $\text{Layer}_{G_1}$ is designed to unfold the following iteration:
\begin{equation}\label{eq.pro_9_G1}
\mathbf{G}_1=\mathbf{G}_1+\eta_1(\mathbf{A}-\mathbf{V}_1).
\end{equation}
With $\mathbf{A}^{(k+1)}$ and $\mathbf{V}_1^{(k+1)}$, the $(k+1)$th neural network layer to update $\mathbf{G}_1$ is designed as
\begin{equation}\label{eq.pro_9_G1_2}
\begin{split}
  \mathbf{G}_1^{(k+1)}&=\text{Layer}_{G_1}\left(\mathbf{A}^{(k+1)},\mathbf{V}_1^{(k+1)};\theta_3^{(k+1)}\right)\\
  &=\mathbf{G}_1^{(k)}+\theta_3^{(k+1)}(\mathbf{A}^{(k+1)}-\mathbf{V}_1^{(k+1)})
\end{split}
\end{equation}
where $\theta_3^{(k+1)}$ is a learnable parameter to replace the role of $\eta_1$.

$\text{Layer}_{G_2}$ is derived from
\begin{equation}\label{eq.pro_9_G2}
\mathbf{G}_2=\mathbf{G}_2+\eta_2(\mathbf{M}-\mathbf{V}_2).
\end{equation}
The same as the design of $\text{Layer}_{G_1}$, this layer uses $\theta_4^{(k+1)}$ to replace the role of $\eta_2$:
\begin{equation}\label{eq.pro_9_G2_2}
\begin{split}
  \mathbf{G}_2^{(k+1)}&=\text{Layer}_{G_2}\left(\mathbf{M}^{(k+1)},\mathbf{V}_2^{(k+1)};\theta_4^{(k+1)}\right)\\
  &=\mathbf{G}_2^{(k)}+\theta_4^{(k+1)}(\mathbf{M}^{(k+1)}-\mathbf{V}_2^{(k+1)}).
\end{split}
\end{equation}

The $\text{Layer}_{G_1}$ and $\text{Layer}_{G_2}$ neural network layers are also shown in Fig.~\ref{fig.unroll}.
\subsection{Network Architecture}
Each unrolling block represents a single iteration of the ADMM based plug-and-play unmixing method and consists of 6 parts: $\text{Layer}_{A}$, $\text{Layer}_{V_1}$, $\text{Layer}_{G_1}$, $\text{Layer}_{M}$, $\text{Layer}_{V_2}$ and $\text{Layer}_{G_2}$. As shown in Fig.~\ref{fig.framework}, we use $K$ iteration blocks to conduct the architecture of our proposed method, which can mimic the ADMM based unmixing algorithm with $K$ iterations. For the $k$th block, the learnable parameters are $\boldsymbol{\Theta}^{(k)}=\left\{\mathbf{R}^{(k)},\mathbf{W}_1^{(k)},\mathbf{H}^{(k)}, \mathbf{Q}_1^{(k)},\theta_1^{(k)},\theta_2^{(k)},\mathbf{W}_2^{(k)}, \mathbf{Q}_2^{(k)},\theta_3^{(k)},\theta_4^{(k)}\right\}$.
Each block in the network has its own set of learnable parameters that are not shared across blocks, which has demonstrated flexibility and strong learning capability for unmixing task.

Additionally, the PnP-Net architecture can be divided into two parts: the endmember estimation network and the abundance estimation network. Each block of the endmember estimation network contains the $\text{Layer}_{M}$, $\text{Layer}_{V_2}$ and $\text{Layer}_{G_2}$ components, there are $K$ blocks to estimate the endmember. The abundance estimation network comprises three parts, namely $\text{Layer}_{A}$, $\text{Layer}_{V_1}$ and $\text{Layer}_{G_1}$. $K$ layers are applied to estimate the abundance.

\subsection{Training Details and Network Initialization}
\label{section:initial}
Our PnP-Net is trained by a blind manner with the input hyperspectral image $\mathbf{X}$. We use the mean square error (MSE) to compute the differences between the input and its reconstruction. 
Compared to calculating the MSE loss directly using the reconstruction of the final $K$th block output, calculating it for each block can better constrain the training process of the network.
Thus, we use the weighted sum of the reconstructions from each block to compute the loss function, which is defined as:
\begin{equation}\label{eq.loss}
  \mathcal{L}=\frac{1}{2N}\sum_{k=1}^{K}\beta_{k}\|\mathbf{X}-\hat{\mathbf{X}}^{(k)}\|_{\text{F}}^2
\end{equation}
where $\beta_{k}$ denotes the importance of each term, $\hat{\mathbf{X}}^{(k)} = \hat{\mathbf{M}}^{(k)}\hat{\mathbf{A}}^{(k)}$ is the corresponding reconstructed image of the $k$th block.

Parameter initialization plays an important role in our PnP-Net, which can accelerate the convergence speed and enhance the endmember and abundance estimation accuracy. The endmember matrix $\mathbf{M}^{(0)}$ and abundance matrix $\mathbf{A}^{(0)}$ are calculated by VCA~\cite{nascimento2005vertex} and FCLS~\cite{heinz2001fully}. Specially, we obtain the initialization of $\mathbf{W}_1$ and $\mathbf{Q}_1$ according to $\mathbf{A}^{(0)}$ and  $\mathbf{M}^{(0)}$, and other values are initialized with zeros.

Our PnP-Net method is flexible to plug various deep denoisers. In this work, we choose 3 novel denoisers, i.e. DnCNN~\cite{zhang2017beyond}, IRCNN~\cite{zhang2017learning} and SCUNet~\cite{zhang2023practical}, as examples. We pretrain these models using images of Waterloo Exploration Database~\cite{ma2016waterloo}, DIV2K~\cite{agustsson2017ntire}, and Flick2K~\cite{liu2018non} datasets with different noise settings. It is worth noting that the parameters of denoiser are fixed during training which can reduce computational stress and take advantage of outer priors. Our PnP-Net with these denoisers are named as PnP-Net-1, PnP-Net-2 and PnP-Net-3. During training, we set the learning rate to $5\times10^{-4}$. The number of iteration blocks ($K$) is set to 5. $L$ and $P$ are set to 3 with the kernel size setting to $1\times1$, $3\times3$ and $5\times5$. We set $\{\beta_k\}_{k=1}^{K}$ to $\left\{1\times 10^{-4},1\times 10^{-3},1\times 10^{-2},1\times 10^{-1},1\right\}$.
\section{Experiment Results}
\label{sec:exp}
In this section, we firstly introduce the evaluation indicators used to analyse the effectiveness of the proposed methods. The considered conventional and state-of-the-art compared methods are also briefly presented. Then, the datasets used in the experiments and results are illustrated and discussed.

\begin{table*}[t]
\centering
\renewcommand\arraystretch{1.25}
\caption{Performance Evaluation of the synthetic data (in terms of aRMSE and PSNR). }\label{Tab_syn_abu_results}
\begin{tabular}{cc|cc|cc|cc|cc}
\hline\hline
\multicolumn{2}{c|}{Noise   level} & \multicolumn{2}{c|}{5dB} & \multicolumn{2}{c|}{10dB} & \multicolumn{2}{c|}{20dB} & \multicolumn{2}{c}{30dB} \\ \hline
\multicolumn{1}{c|}{Method} & Denoiser & \multicolumn{1}{c|}{aRMSE~$\downarrow$} & PSNR~$\uparrow$ & \multicolumn{1}{c|}{aRMSE~$\downarrow$} & PSNR~$\uparrow$& \multicolumn{1}{c|}{aRMSE~$\downarrow$} & PSNR~$\uparrow$& \multicolumn{1}{c|}{aRMSE~$\downarrow$} & PSNR~$\uparrow$\\ \hline
\multicolumn{1}{c|}{SUnSAL-TV} & / & \multicolumn{1}{c|}{0.0991} & 27.8650 & \multicolumn{1}{c|}{0.0658} & 30.5224 & \multicolumn{1}{c|}{0.0225} & 44.0060 & \multicolumn{1}{c|}{0.0081} & 50.5379 \\ \hline
\multicolumn{1}{c|}{gtvMBO} & / & \multicolumn{1}{c|}{0.1149} & 29.6094 & \multicolumn{1}{c|}{0.0699} & 34.3044 & \multicolumn{1}{c|}{0.0272} & 43.9903 & \multicolumn{1}{c|}{{\ul 0.0080}} & 53.8601 \\ \hline
\multicolumn{1}{c|}{U-ADMM-BUNet} & / & \multicolumn{1}{c|}{0.0939} & 30.6719 & \multicolumn{1}{c|}{0.0647} & 35.1446 & \multicolumn{1}{c|}{0.0256} & 44.0009 & \multicolumn{1}{c|}{0.0086} & 53.3713 \\ \hline
\multicolumn{1}{c|}{DIFCNN} & / & \multicolumn{1}{c|}{0.1044} & 30.4841 & \multicolumn{1}{c|}{0.0651} & 34.8753 & \multicolumn{1}{c|}{0.0242} & 44.0677 & \multicolumn{1}{c|}{0.0081} & 53.1629 \\ \hline
\multicolumn{1}{c|}{\multirow{2}{*}{AERED}} & NLM & \multicolumn{1}{c|}{0.0907} & 32.2215 & \multicolumn{1}{c|}{0.0604} & 35.6194 & \multicolumn{1}{c|}{0.0246} & 44.1319 & \multicolumn{1}{c|}{0.0081} & 54.4845 \\ 
\multicolumn{1}{c|}{} & BM3D & \multicolumn{1}{c|}{0.0927} & 32.4679 & \multicolumn{1}{c|}{0.0597} & 36.3611 & \multicolumn{1}{c|}{0.0217} & 44.1408 & \multicolumn{1}{c|}{{\ul 0.0080}} & 54.6312 \\ \hline
\multicolumn{1}{c|}{\multirow{3}{*}{PnP-Net}} & DnCNN & \multicolumn{1}{c|}{{\ul 0.0856}} & \textbf{33.5082} & \multicolumn{1}{c|}{0.0572} & 37.6063 & \multicolumn{1}{c|}{\textbf{0.0197}} & \textbf{44.3506} & \multicolumn{1}{c|}{\textbf{0.0079}} & \textbf{55.5990} \\ 
\multicolumn{1}{c|}{} & IRCNN & \multicolumn{1}{c|}{\textbf{0.0850}} & {\ul 33.4259} & \multicolumn{1}{c|}{{\ul 0.0566}} & \textbf{37.7314} & \multicolumn{1}{c|}{0.0203} & 44.1599 & \multicolumn{1}{c|}{{\ul 0.0080}} & 55.4729 \\ 
\multicolumn{1}{c|}{} & SCUNet & \multicolumn{1}{c|}{0.0859} & 33.3130 & \multicolumn{1}{c|}{\textbf{0.0535}} & {\ul 37.7282} & \multicolumn{1}{c|}{{\ul 0.0199}} & {\ul 44.2947} & \multicolumn{1}{c|}{{\ul 0.0080}} & {\ul 55.4810} \\ \hline\hline
\end{tabular}
\end{table*}

\begin{table*}[t]
\centering
\renewcommand\arraystretch{1.25}
\caption{Performance Evaluation of the synthetic data (in terms of mRMSE and mSAD). }\label{Tab_syn_end_results}
\begin{tabular}{cc|cc|cc|cc|cc}
\hline\hline
\multicolumn{2}{c|}{Noise   level} & \multicolumn{2}{c|}{5dB} & \multicolumn{2}{c|}{10dB} & \multicolumn{2}{c|}{20dB} & \multicolumn{2}{c}{30dB} \\ \hline
\multicolumn{1}{c|}{Method} & Denoiser & \multicolumn{1}{c|}{mRMSE~$\downarrow$} & mSAD~$\downarrow$ & \multicolumn{1}{c|}{mRMSE~$\downarrow$} & mSAD~$\downarrow$ & \multicolumn{1}{c|}{mRMSE~$\downarrow$} & mSAD~$\downarrow$ & \multicolumn{1}{c|}{mRMSE~$\downarrow$} & mSAD~$\downarrow$ \\ \hline
\multicolumn{1}{c|}{SUnSAL-TV} & / & \multicolumn{1}{c|}{0.0402} & 6.3464 & \multicolumn{1}{c|}{0.0228} & 3.6377 & \multicolumn{1}{c|}{0.0052} & 0.9584 & \multicolumn{1}{c|}{0.0023} & 0.2176 \\ \hline
\multicolumn{1}{c|}{gtvMBO} & / & \multicolumn{1}{c|}{0.0502} & 6.4056 & \multicolumn{1}{c|}{0.0205} & 2.7845 & \multicolumn{1}{c|}{0.0100} & 1.1000 & \multicolumn{1}{c|}{{\ul 0.0014}} & \textbf{0.1913} \\ \hline
\multicolumn{1}{c|}{U-ADMM-BUNet} & / & \multicolumn{1}{c|}{0.0400} & 6.3073 & \multicolumn{1}{c|}{0.0200} & 3.2512 & \multicolumn{1}{c|}{0.0050} & 0.9482 & \multicolumn{1}{c|}{0.0015} & 0.2003 \\ \hline
\multicolumn{1}{c|}{DIFCNN} & / & \multicolumn{1}{c|}{0.0423} & 6.4225 & \multicolumn{1}{c|}{0.0199} & 3.0286 & \multicolumn{1}{c|}{0.0067} & 1.1064 & \multicolumn{1}{c|}{0.0017} & 0.2001 \\ \hline
\multicolumn{1}{c|}{\multirow{2}{*}{AERED}} & NLM & \multicolumn{1}{c|}{0.0415} & 6.2650 & \multicolumn{1}{c|}{0.0204} & 2.9087 & \multicolumn{1}{c|}{0.0051} & 1.0135 & \multicolumn{1}{c|}{0.0015} & 0.2088 \\ 
\multicolumn{1}{c|}{} & BM3D & \multicolumn{1}{c|}{0.0401} & 6.2937 & \multicolumn{1}{c|}{0.0206} & 2.7587 & \multicolumn{1}{c|}{0.0049} & 0.9676 & \multicolumn{1}{c|}{0.0014} & 0.2069 \\ \hline
\multicolumn{1}{c|}{\multirow{3}{*}{PnP-Net}} & DnCNN & \multicolumn{1}{c|}{{\ul 0.0394}} & {\ul 6.2195} & \multicolumn{1}{c|}{{\ul 0.0194}} & {\ul 2.7191} & \multicolumn{1}{c|}{{\ul 0.0046}} & \textbf{0.8658} & \multicolumn{1}{c|}{\textbf{0.0013}} & 0.1928 \\ 
\multicolumn{1}{c|}{} & IRCNN & \multicolumn{1}{c|}{0.0399} & 6.2858 & \multicolumn{1}{c|}{0.0202} & 2.7848 & \multicolumn{1}{c|}{0.0049} & 0.9291 & \multicolumn{1}{c|}{{\ul 0.0014}} & 0.1943 \\ 
\multicolumn{1}{c|}{} & SCUNet & \multicolumn{1}{c|}{\textbf{0.0390}} & \textbf{6.1300} & \multicolumn{1}{c|}{\textbf{0.0192}} & \textbf{2.6464} & \multicolumn{1}{c|}{\textbf{0.0042}} & {\ul 0.8734} & \multicolumn{1}{c|}{{\ul 0.0014}} & {\ul 0.1941} \\ \hline\hline
\end{tabular}
\end{table*}

\subsection{Evaluation Indicators}
We use several indicators to evaluate the unmixing results quantitatively in terms of estimated abundances, extracted endmembers and reconstructed images. In addition, we have no available ground truth for the real datasets, we only evaluate the reconstructed image quantitatively and subjectively evaluate the visual effect.

The root mean square error (RMSE) is applied to assess the differences between the ground truth and estimated abundances and endmembers. We respectively name them as aRMSE and mRMSE, which are defined as follows:
\begin{equation}
\mathrm{aRMSE}=\sqrt{\frac{1}{N R} \sum_{i=1}^N\left\|\mathbf{a}_i-\hat{\mathbf{a}}_i\right\|^2}
\end{equation}
and
\begin{equation}
\mathrm{mRMSE}=\sqrt{\frac{1}{B R} \sum_{r=1}^R\left\|\mathbf{m}_r-\hat{\mathbf{m}}_r\right\|^2},
\end{equation}
where $\hat{\mathbf{a}}_i$ and $\mathbf{a}_i$ represent the estimated and real abundances of the $i$th pixel, and $\hat{\mathbf{m}}_r$ and $\mathbf{m}_r$ denote the $r$th extracted and corresponding real endmembers.

We apply the mean spectral angle distance (SAD) to evaluate the degree of distortion between the real and reconstructed spectral signatures, and mSAD denotes the SAD between the estimated and real endmembers. They are calculated by:
\begin{equation}
\mathrm{SAD}=\frac{1}{N} \sum_{i=1}^N \arccos \left(\frac{\mathbf{y}_i^{\top} \hat{\mathbf{y}}_i}{\left\|\mathbf{y}_i\right\|\left\|\hat{\mathbf{y}}_i\right\|}\right)
\end{equation}
and
\begin{equation}
\mathrm{mSAD}=\frac{1}{R} \sum_{r=1}^R \arccos \left(\frac{\mathbf{m}_i^{\top} \hat{\mathbf{m}}_r}{\left\|\mathbf{m}_r\right\|\left\|\hat{\mathbf{m}}_r\right\|}\right)
\end{equation}
where $\mathbf{y}_i$ and $\hat{\mathbf{y}}_i$ represent the real and reconstructed spectra.

We also use the peak signal-to-noise ratio (PSNR) to evaluate the differences between the real and reconstructed images, calculated by:
\begin{equation}
\mathrm{PSNR}=10 \times \log _{10}\left(\frac{\mathrm{max}^2}{\mathrm{MSE}}\right)
\end{equation}
where $\mathrm{max}$ represents the highest pixel value in the reconstructed image.
\begin{figure*}[t]
  \centering
  \includegraphics[width=18cm]{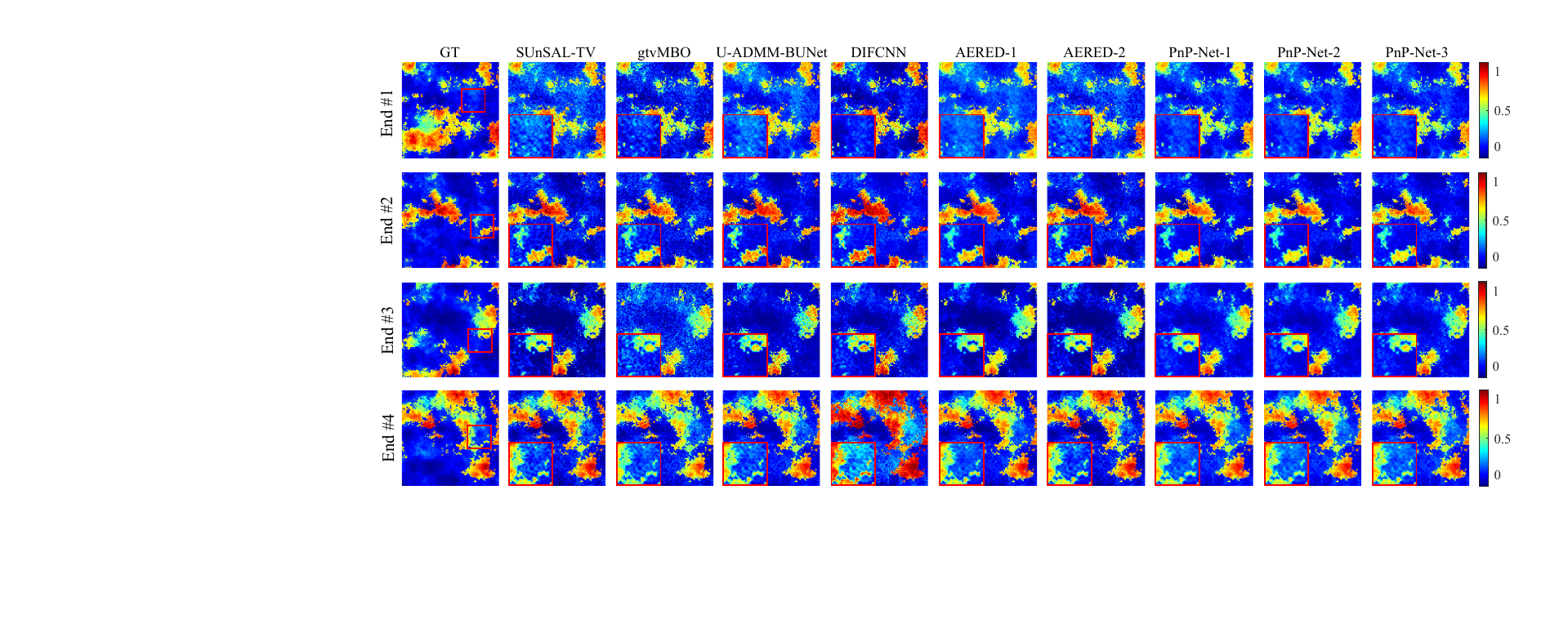}\\
  \caption{Illustration of the ground-truth and estimated abundance maps (SNR = 10dB). From top to bottom different endmembers. From left to right different compared methods. The subfigures zoom out the region within the red box on the ground-truth map.}\label{fig.abu_syn}
\end{figure*}

\begin{figure*}[t]
  \centering
  \includegraphics[width=18cm]{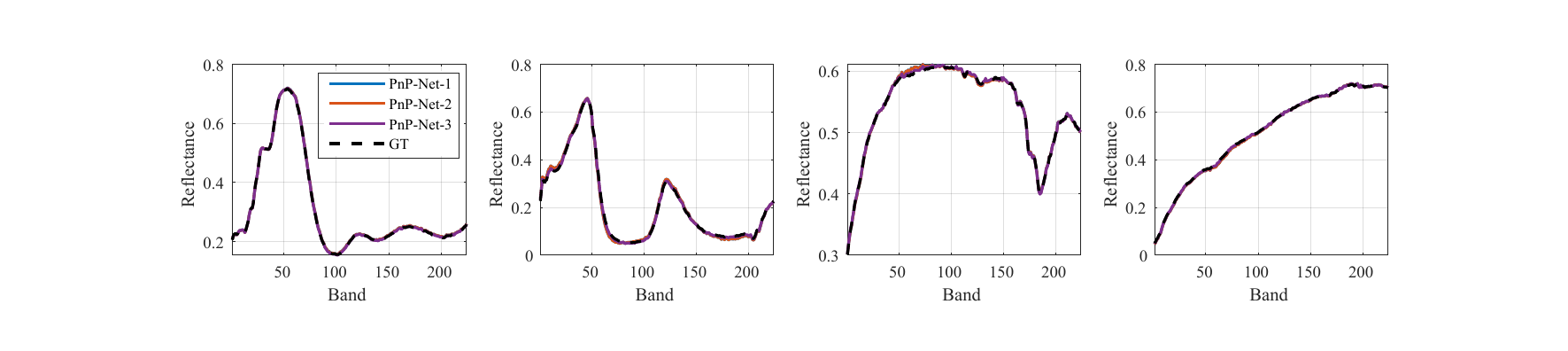}\\
  \caption{Illustration of the ground-truth and estimated endmembers (SNR = 20dB).}\label{fig.end_syn}
\end{figure*}
\begin{figure}
  \centering
  \includegraphics[width=9cm]{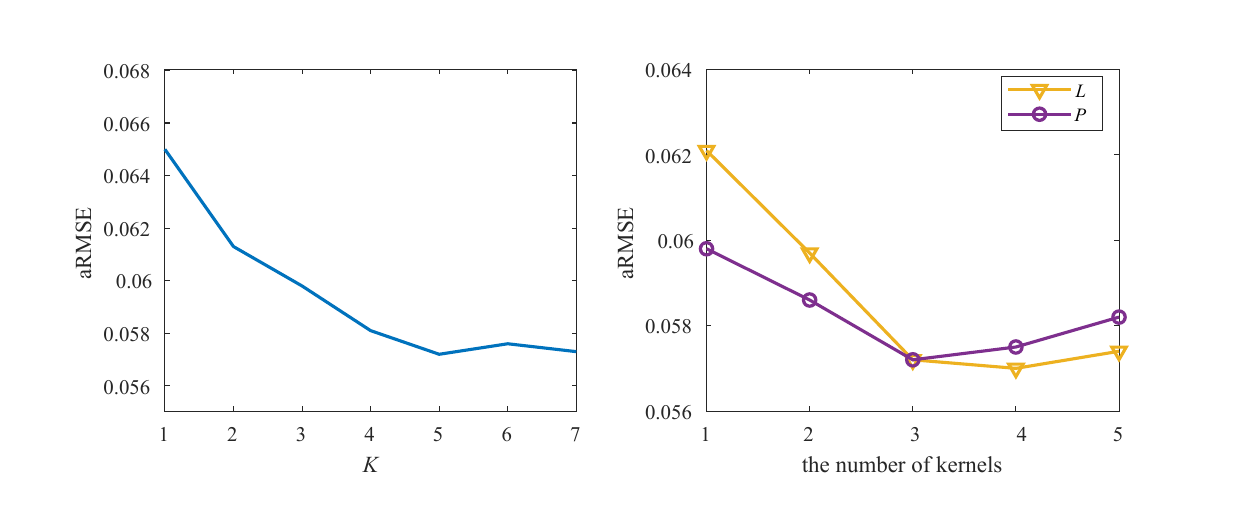}\\
  \caption{The left subfigure shows the impact of the number of iteration blocks $K$ on abundance estimation, and the right subfigure shows the impact of the number of kernels $L$ and $P$ on abundance estimation.}\label{fig.para}
\end{figure}
\subsection{Compared Methods}
We compare the proposed PnP-Net method with several conventional and state-of-the-art unmixing methods, i.e. SUnSAL-TV~\cite{iordache2012total}, gtvMBO~\cite{qin2020blind}, U-ADMM-BUNet~\cite{zhou2021admm}, DIFCNN~\cite{kong2023deep} and AERED~\cite{zhao2024ae}. To make a fair comparison, all methods are initialized by the endmembers extracted by VCA~\cite{nascimento2005vertex} and abundances estimated by FCLS~\cite{heinz2001fully}.  All parameters of these methods are carefully chosen to achieve the best experimental results.

The first two compared methods are conventional methods. The SUnSAL-TV method uses the VCA to extract the endmembers, and a total variation spatial regularization is applied to constraint the estimated abundances. The gtvMBO method is a blind unmixing method with a data-driven graph total variation regularization, and its objective function is solved by ADMM. The U-ADMM-BUNet and DIFCNN are novel unrolling based unmixing methods. 
AERED integrates the autoencoder based unmixing network with RED, which combines the explicit and implicit priors. We name the AERED with NLM denoiser and block-matching and
4-D ffltering (BM4D)~\cite{maggioni2012nonlocal} deoiser as AERED-1 and AERED-2, respectively.

\begin{figure*}[t]
  \centering
  \includegraphics[width=17cm]{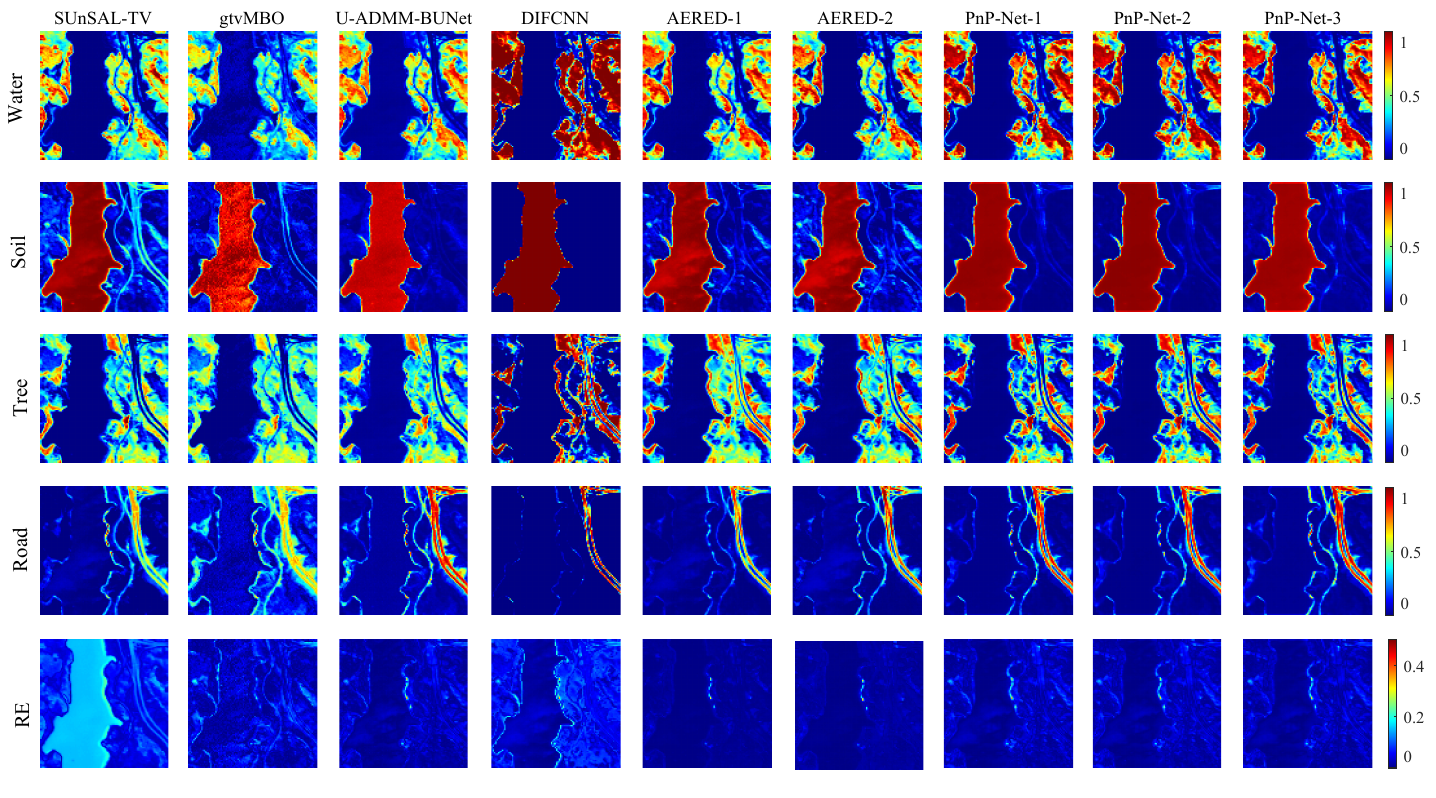}\\
  \caption{Illustration of the results of compared methods and proposed PnP-Net on Jasper Ridge dataset. (The first four rows are estimated abundances of different endmembers, and the last row is the map of reconstructed error.) }\label{fig.abu_jas}
\end{figure*}

\begin{figure*}[t]
  \centering
  \includegraphics[width=17cm]{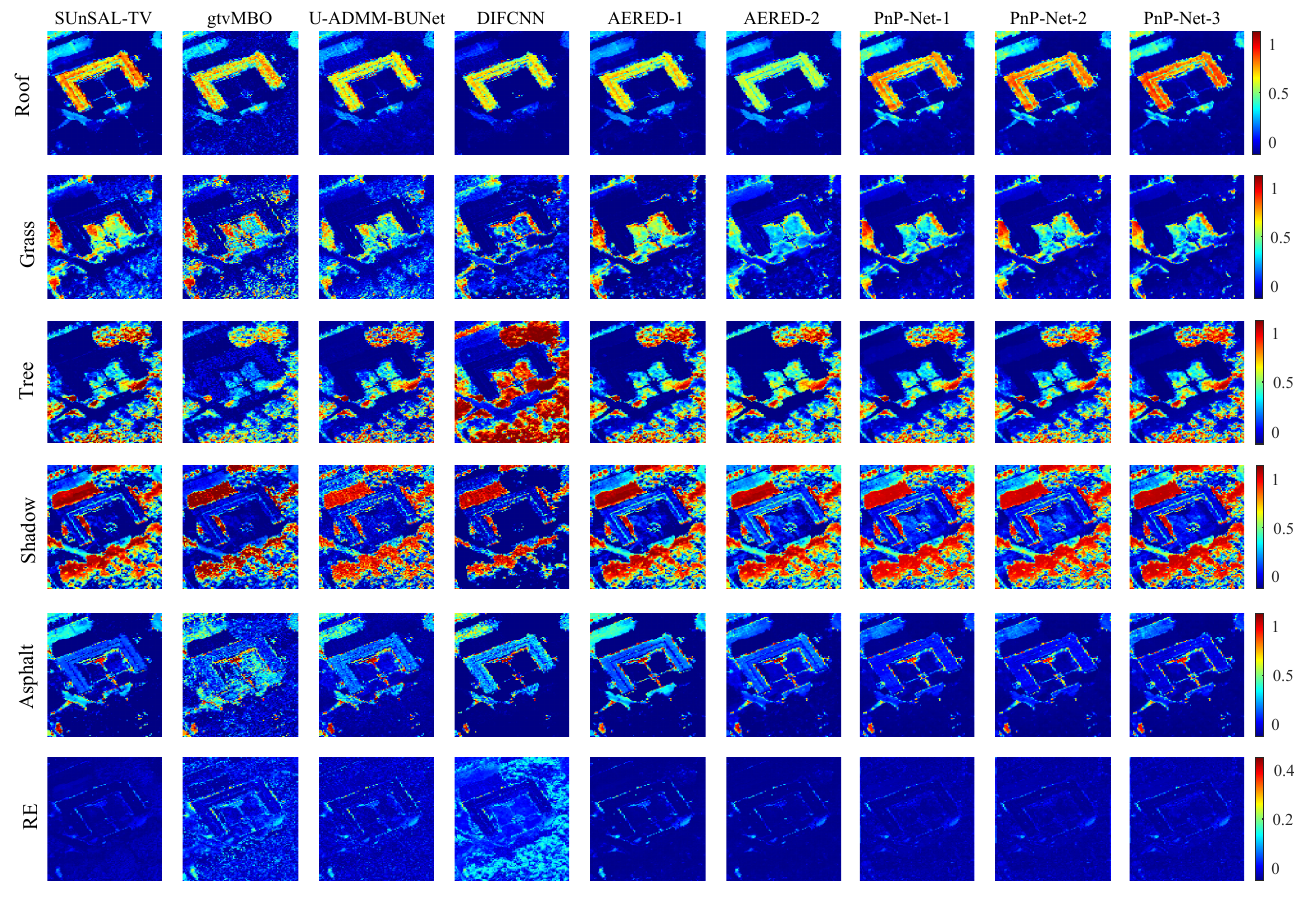}\\
  \caption{Illustration of the results of compared methods and our PnP-Net on Muufl Gulfport dataset. (The first five rows are estimated abundances of different endmembers, and the last row is the map of reconstructed error.)}\label{fig.abu_muffle}
\end{figure*}
\subsection{Experiments on Synthetic Datasets}
In order to quantitatively verify the unmixing results, we generate a set of synthetic data with the spatial size of $100\times 100$. We adapt the method described in~\cite{han2022autonas} to produce these data. The ground-truth of abundance maps are shown in Fig.~\ref{fig.abu_syn}, and they are constrained to satisfy ASC and ANC. The abundance obtained by this strategy shows spatial correlation that similar to the real data. Four spectral signatures selected from the USGS spectral library with 224 bands are applied to generate the endmember matrix. The reference endmember spectra are shown in Fig.~\ref{fig.end_syn}. Zero mean Gaussian with four levels of signal-to-noise-ratio (SNR), i.e. 5dB, 10dB, 20dB and 30dB, are added to the data.

Table~\ref{Tab_syn_abu_results} shows the aRMSE and PSNR results of the synthetic data, and Table~\ref{Tab_syn_end_results} lists the mRMSE and mSAD results.
The bold numbers in Tables~\ref{Tab_syn_abu_results} and~\ref{Tab_syn_end_results} denote the best results, and the underlined numbers indicate the second best results.
From the results of these two tables, it can be observed that the proposed PnP-Net unmixing framework achieves the best unmixing results compared to the benchmark methods. Compared to traditional unmixing methods such as SUnSAL-TV and gtvMBO, the proposed method eliminates the need for penalty parameter selection, instead utilizing a data-driven learning approach. In comparison to unrolling unmixing method, i.e. U-ADMM-BUNet and DIFCNN, this proposed PnP-Net method employs dynamic convolution to achieve multiscale information learning and fusion and gets better results. Our PnP-Net method leverages an externally pretrained denoiser on large datasets to provide outer priors, and also integrate a hyperspectral data-driven and physically interpretable neural network to learn internal priors. This combination results in superior unmixing performance. We can also observe that, due to the use of deep denoisers and deep neural networks, our proposed unmixing method is robust to noise.

Fig.~\ref{fig.abu_syn} shows the abundance maps of the proposed method and the comparison methods for the synthetic data at 10dB. We observe that all these estimates are very close to the ground-truth. Moreover, it can be seen that the noise in the abundance maps of PnP-Net and AERED is obviously smaller than that of other comparison methods, which indicates the superiority of using denoiser to bring prior information. Furthermore, Fig.~\ref{fig.end_syn} illustrates the extracted endmembers of the proposed method. We observe that the result of our method is close to the ground-truth.
Fig.~\ref{fig.para} illustrates how the number of blocks affect the unmixing performance and also shows that how the
number of kernels $L$ and $P$ impact the unmixing results.
\subsection{Experiments on Real Datasets}
We also evaluate our proposed method on real datasets. We conduct experiments on two widely used hyperspectral images obtained by airborne sensors, namely Jasper Ridge dataset and Muufl Gulfport dataset. The descriptions and experimental results about these two datasets are presented below.

The Jasper Ridge dataset is captured by Analytical Imaging and Geophysics (AIG) in 1999. It contains 224 bands covering the spectral range from 380nm to 2500nm. After removing the bands affected by water vapor and atmospheric (1-3, 108-112, 154-166, and 220-224), 198 bands are remained. The original size of this data is $512\times 614$, a popular region of interest with $100\times 100$ pixels are cropped. The four endmembers in this data are ``water", ``soil", ``tree" and ``road".

The Muufl Gulfport dataset was captured by the CASI-1500 hyperspectral sensor over the University of Southern Mississippi Gulf Park Campus. The original image contains 72 bands with $325\times 220$ pixels. We remove the first four and the last four bands of the image, which contain a lot of noise. A subimage with the size of $130\times 90$ pixels is applied in our experiments. We reference the scene label ground-truth map from manually labeling in~\cite{du2017technical}. There are 5 pure materials in this subimage, namely ``roof", ``grass", ``tree", ``shadow'' and ``asphalt".

The visual results of the estimated abundance maps of these two datasets are shown in Fig.~\ref{fig.abu_jas} and Fig.~\ref{fig.abu_muffle}.
The first four rows of Fig.~\ref{fig.abu_jas} report the comparisons of our method and compared methods with respect to the Jasper Ridge dataset. All these methods decompose this data with four clear abundance maps. But we observe that the abundance maps of ``road" material of SUnSAL-TV and gtvMBO is lighter than other methods. The reason may be that the data contains spectral variability, which makes these two conventional methods hard to extract representative endmember. Our PnP-Net gets clearer abundance results. The last line of Fig.~\ref{fig.abu_jas} illustrates the reconstructed error between the reconstructed image and real image of these methods. We observe that our method gets good reconstruct results which are also consistent with the results in Table~\ref{Tab_real_jas}. To some extent, it proves the advantages of our method to unmix the real data. The abundance maps of Muufl Gulfport dataset are illustrated in
Fig.~\ref{fig.abu_muffle}. It can be seen that the results of gtvMBO are noisy than other methods, especially for the ``asphalt" endmember. The proposed PnP-Net method can estimate the abundance maps much clearer than the compared method. Our results also have more detailed spatial information and sharper edges. The last row of Fig.~\ref{fig.abu_muffle} presents the reconstructed error maps of all methods. DIFCNN shows higher results. It may cause by the loss function selected to train the model. It use the cross entropy-loss to train the network, while other compared deep methods apply the MSE-like loss to train the model. The latter is more conductive to reduce the Euclidean distance between the reconstructed and original image. Table~\ref{Tab_real_mul} lists quantitative results of the Muufl Gulfport data. We get that our method can obtain good reconstructions.
\begin{table*}[t]
\centering
\renewcommand\arraystretch{1.25}
\caption{Reconstruction Performance Evaluation of the Jasper Ridge data (in terms of PSNR and SAD). }\label{Tab_real_jas}
\begin{tabular}{c|c|c|c|c|cc|ccc}
\hline\hline
Method & SUnSAL-TV & gtvMBO & U-ADMM-BUNet & DIFCNN & \multicolumn{2}{c|}{AERED} & \multicolumn{3}{c}{PnP-Net} \\ \hline
Denoiser & / & / & / & / & \multicolumn{1}{c|}{NLM} & BM3D & \multicolumn{1}{c|}{DnCNN} & \multicolumn{1}{c|}{IRCNN} & SCUNet \\ \hline
PSNR~$\uparrow$ & 28.5721 & 29.5939 & 30.0075 & 29.7343 & \multicolumn{1}{c|}{\textbf{31.3895}} & {30.0416} & \multicolumn{1}{c|}{{29.9136}} & \multicolumn{1}{c|}{{\ul 30.1203}} & 30.0763 \\ \hline
SAD~$\downarrow$ & 9.2143 & 6.3181 & 5.6415 & 8.7582 & \multicolumn{1}{c|}{5.5519} & {\ul 5.4386} & \multicolumn{1}{c|}{5.5369} & \multicolumn{1}{c|}{\textbf{5.4210}} & 5.6177 \\ \hline\hline
\end{tabular}
\end{table*}
\begin{table*}[t]
\centering
\renewcommand\arraystretch{1.25}
\caption{Reconstruction Performance Evaluation of the Muufl Gulfport data (in terms of PSNR and mSAD). }\label{Tab_real_mul}
\begin{tabular}{c|c|c|c|c|cc|ccc}
\hline\hline
Method & SUnSAL-TV & gtvMBO & U-ADMM-BUNet & DIFCNN & \multicolumn{2}{c|}{AERED} & \multicolumn{3}{c}{PnP-Net} \\ \hline
Denoiser & / & / & / & / & \multicolumn{1}{c|}{NLM} & BM3D & \multicolumn{1}{c|}{DnCNN} & \multicolumn{1}{c|}{IRCNN} & SCUNet \\ \hline
PSNR~$\uparrow$ & 31.5959 & 30.1698 & 31.1459 & 25.652 & \multicolumn{1}{c|}{32.3154} & \textbf{33.4423} & \multicolumn{1}{c|}{32.4775} & \multicolumn{1}{c|}{{\ul 32.5454}} & 32.3401 \\ \hline
SAD~$\downarrow$ & 5.2867 & 7.5754 & 6.8583 & 8.3911 & \multicolumn{1}{c|}{5.1006} & \textbf{4.4823} & \multicolumn{1}{c|}{4.9912} & \multicolumn{1}{c|}{{\ul 4.9505}} & 5.1039 \\ \hline\hline
\end{tabular}
\end{table*}
\section{Conclusion}
\label{sec:conclusion}
In this paper, we propose a novel PnP-Net method for hyperspectral unmixing. We unroll the plug-and-play unmixing method into trainable deep model. We apply the RED to add denoising priors, and a various of denoisers can be plugged. The ADMM is used to solve the optimization function. We unfold the ADMM based iterative steps to design the layer, and one layer represents one iteration. $K$ layers are stacked to conduct the deep architecture. Through unrolling, we can use the end-to-end manner to train the network. To fully exploit the spatial information, we use the dynamic convolution with different sizes of convolution kernels to capture multiscale information. The denoisers are pretrained with a variety of images which make our method leverage the power of external information to enhance the modeling ability. This strategy bypasses the issue of limited hyperspectral data, and provides a way to integrate the inner and outer priors. Extensive experiment results show the superior performance of our method.
\bibliographystyle{IEEEtran}
\bibliography{IEEEfull,BIB}\ 

\end{document}